\documentclass[12pt]{article}
\usepackage{amssymb, color, epsfig, amsmath, pifont, natbib, enumitem, graphicx,mathrsfs}
\usepackage{fancyhdr,appendix}
\graphicspath{{./}{../graphs/}}
\makeatletter
\renewcommand*\@fnsymbol[1]{\the#1}
\makeatother

\title{Towards a Sustainable Agricultural Credit Guarantee Scheme}
\author{ Reason L. Machete\footnote{Email: rmachete@bitri.co.bw, machete.r.l@gmail.com, r.l.machete@lse.ac.uk; Tel: +267-77610677.}\\\\
{\footnotesize  Climate Change, Botswana Institute for Technology Research and Innovation, Botswana}\\
{\footnotesize  Statistics Department, London School of Economics and Political Science, United Kingdom}}
\date{August 5, 2020}
\begin{document}
\maketitle
\noindent\rule{14.9cm}{0.4pt}
\begin{abstract}
Since 1986, Government of Botswana has been running an Agricultural Credit Guarantee Scheme for dry-land arable farming. The scheme purports to assist dry-land crop farmers who have taken loans with participating banks or lending institutions to help them meet their debt obligations in case of crop failure due to drought, floods, frost or hailstorm. Nonetheless, to date, the scheme has focused solely on drought. The scheme has placed an unsustainable financial burden on Government because it is not based on sound actuarial principles. This paper argues that the level of Government subsidies should take into account the gains made by farmers during non-drought years. It is an attempt to circumvent the challenges of correlated climate risks and recommends a quasi self-financing mechanism, assuming that the major driver of crop yield failure is drought. Moreover, it provides a novel subsidy and premium rate setting method.
\end{abstract}
\noindent\rule{14.9cm}{0.4pt}\\
\textsf{\copyright} 2019, Machete
\section{Introduction}
In many countries around the world, the importance of agricultural produce is underscored by the efforts to establish insurance for this specific sector of the economy. The USA, Germany, Italy, France and Spain are examples of countries that have established agricultural insurance~\citep{dick-2010}. Agricultural insurance in these countries depends heavily on Government subsidies. The subsidies tend to be channelled towards premiums. Among these countries, Spain is said to have the most developed system of agricultural insurance that covers several risks~\citep{col-14}. In essence it provides a full-risk coverage. It is noteworthy that increasing the number of perils covered will, inevitably, increase the premiums.

Let it be noted, however, that the term agricultural is broad and includes many perils that affect crops and livestock such as hail, floods, frost, etc. Drought is but just one of these many perils that are grouped together under agricultural insurance. When drought is frequent as is the case in the US, Spain and China, it is difficult, if not impossible, to insure without Government subsidy. Actually, agricultural insurance in developing countries is mostly subsidised~\citep{alam-2020}. There is a sizable body of literature on Government subsidised multi-peril crop insurance in the USA~\citep[e.g.][]{du-2017,dun-00,glauber-2004,ker-2000} and most of it is not solely focused on drought insurance for crops. Some literature talks of yield insurance, which naturally includes drought insurance~\citep[e.g.][]{eze-2020,wang-03}. In this paper we consider yield to be a good proxy for drought and use it to make an empirical assessment of drought in Botswana. It is preferred to rainfall because our analysis has shown rainfall to be a poor index of crop yield. While there are other possible indices to consider, these also fail to capture crop yield for dry land crops~\citep{reeves-17}. In particular, it is difficult to detect any correlations between yield for dry land crops and any of the indices. Such problems can lead to what is referred to as {\em basis risk}. Basis risk refers to a case where the payout is not sufficient to defray the losses incurred by the insured~\citep{eze-2020}.
%\section{Drought in Botswana}

Meanwhile, Botswana has had an Agricultural Credit Guarantee Scheme to manage the {\it agricultural} drought that is specific to dry-land crops since 1986. Administered by Government, this is supposed to be a multi peril insurance even though, since its inception, it has solely covered drought following presidential declarations. Every year, at the end of the harvest, a team of experts from various Government departments travels around the country to perform drought assessment and subsequently make recommendations to Government. The team is referred to as an Inter-Ministerial Drought Assessment Committee. Following the team's recommendations, a national or regional drought may be declared with relevant interventions to be implemented. The interventions entail two aspects: (i) Drought Relief Interventions and (ii) Contributions to instalments for loans taken by arable farmers under the Agricultural Credit Guarantee Scheme. These interventions are performed only if a presidential declaration of drought has been made. Drought interventions can include supplementary feeding for children under 5 years old, employment of casual workers, construction projects, seed subsidy for farmers, subsidy on selected cattle feeds and maintenance of fire breaks. In addition to being given loan subsidies when they have taken loans within participating banks, subsidy for commercial farmers is limited to Government contributing 65\% towards the cost of fertilisers, herbicides and pesticides. Several years, the Ministry of Agriculture in Botswana commissioned a consultancy to investigate the possibility of supporting farming through an agricultural insurance sscheme~\cite{ker-07a}, but this did not bear significant fruit because it lacked concrete recommendations of how to setup such a scheme.

\begin{table}
\begin{center}
\begin{tabular}{|c|c|c|c|c|c|}
\hline
   SEASON  & 2004/5 & 2005/6 & 2006/7 & 2014/15 & 2015/16 \\
     \hline\hline
     DRI & P 205 M & & P 324 M & P 77 M & P 632 M\\
     \hline
     ACGS & P 17 M & P 22 M & P 8 M & P 244 M & P 115 M\\
     \hline
     TOTAL & P 222 M & P 22 M & P 332 M & P 321 M & P 747 M \\
     \hline
\end{tabular}
\end{center}
\caption{\em\small A sample of Government expenses following the drought seasons indicated. DRI denotes Drought Relief Interventions and ACGS denotes Agricultural Credit Guarantee Scheme. The ACGS amounts indicated were taken only through the National Development Bank. The amounts taken through CEDA were not supplied.}
\label{tab:paid}
\end{table}
A sample of Government expenses following drought declarations is shown in Table~\ref{tab:paid}. A number of observations can be made based on past drought declarations and subsequent payouts. Most importantly, an analysis of the data indicates that, on average, drought is declared about twice in every three years and is thus a high-frequency event. It has been argued that insurance is not the right mechanism for such high frequency events~\citep[e.g.][]{clarke-2013}. Some of the droughts can be quite extreme, such as that of the 2015/2016 season. It is worth noting that such an extreme drought is a one in 35 year event. Irrespective of the impact of the drought, we argue here that the Agricultural Credit Guarantee Scheme is not sustainable because it is not based on actuarial sound principles. In essense, Government may have irrationally committed itself to highly subsidise dry land arable farming. The level of cover it provides is inconsistent with contributions to the scheme. This paper further argues that contributions should be specific to the crops planted in a given season rather than just remain generic. It provides novel tools that can be used to determine the level of Government subsidy necessary to keep farmers solvent.
\section{How Insurance Works}
It has been highlighted in the previous section that drought declarations are made 2/3 of the time and, after each declaration, Government spends millions of Pula on drought relief interventions. A simple average of data on Table~\ref{tab:paid} suggests that an average of around P322 million per annum is spent by Government on drought interventions. This is surely a lot of money that Government spends every year. It should be understood, however, that any insurance company should charge not less than the Government's average loss per annum. Otherwise the insurer cannot remain solvent. In fact, the insurer should charge more than the Government's average loss to take into account administrative costs of running the insurance. 

What then is the value of insurance and how does it work? Insurance derives its value from keeping expenses (in the form of premiums) fixed from year to year whilst the insurance company takes the risk of fluctuating payouts. As defined by~\cite{mehr-85}, insurance is an instrument for reducing risk by combining a sufficient number of exposure units to make their individual losses collectively predictable. Combining several exposure units is called {\emph risk pooling}. What is the benefit of risk pooling then? An insurer benefits from risk pooling in the sense that {\em the law of large numbers} insures that fluctuations are reduced when the insured entities increase. Furthermore, the insured can benefit from the law of large numbers through reduced administrative and buffer load costs due to cost sharing~\citep{wang-03}. The benefit of risk pooling is not loss reduction but risk reduction! Whereas, traditionally, the insured units have been required to be uncorrelated~\citep[e.g. see][]{brown-07}, it has been shown that risk pooling can still reduce risk even when the insured units are imperfectly correlated~\citep{annan-2013,wang-03}.

Consider an insurer who receives gross premium $P$. The gross premium is the sum
\begin{align}
P=P_n+L+A,
\end{align}
where $P_n$ is the net premium, $L$ is the buffer load and $A$ is the term for administrative costs~\citep{wang-03}. The buffer load term takes care of variations in claims frequency and severity and it diminishes to zero as the pool gets larger. Each insured contributes an amount equivalent to their average loss to the net premium and some additional amount to the buffer load and administrative costs~\citep{brown-07}. It is important to note that the insured benefit from risk pooling through a reduction in their contributions towards the buffer load and administrative costs. Risk pooling will not and cannot reduce their individual contributions towards the net premium!

It might seem to some that the case of Botswana's drought is markedly different from the car insurance industry. In the case of car insurance, it would appear that one contributes low premiums relative to the size of possible damages for which they are covered. In fact, in Botswana, insurers typically charge premiums not more than 5\% of the value of the car. Such a low percentage could be the one that inspired the 5\% premiums charged under the Agricultural Credit Guarantee Scheme. A fraction consistent with the drought situation of Botswana should, however, not be less than the horrifying 2/3 (or 66.67\%). In contrast, why do insurers charge low percentage premiums for car insurance? The reason is that the low percentages are consistent with the risk that is underwritten. In particular, one should imagine that the risk of a car being damaged beyond economic value is below 5\% on average. In order to appreciate this percentage, think of how rare it is that one gets involved in a serious accident. Perhaps in your lifetime you have had no more than one or two serious accidents. Most of the minor ones can typically be settled by {\em excess}\footnote{Simply put, {\em excess} is the amount below which an insurance company does not issue a payout. In actuarial language, this is called a {\em deductible}~\citep{brown-07}.}, perhaps avoiding the fines incurred when incidents  are reported to the police. Bearing the foregoing issues in mind, can we come up with a contribution plan that is based on sound actuarial principles?
\section{An Alternative Contribution Plan}
From the past records, it could not be established which declarations were regional and which were national. Therefore, we will assume that all declarations that resulted in payouts were national because an overwhelming majority of declarations appear to have been national. This assumption also aids tractability of the problem. We argue that contributions should be specific to the crops planted in a given planting season and consistent with the respective area planted. If there are $J$ types of crops, we can let $j$ be the index for a given crop, with $j\in\{1,\ldots,J\}$. In the subsequent discussions, $\omega$ will be taken to be the frequency of drought declarations. Taking $\mathit{\Upsilon}_j(t)$ to be the yield for the $j^{\mbox{th}}$ crop at time $t$, we can define the corresponding crop-specific drought threshold to be $\mu_c^{(j)}$ such that 
\begin{equation}
F_j\left(\mu_c^{(j)}\right)=\omega,
\end{equation}
where $F_j(\cdot)$ is the cumulative distribution function of the yield for the $j^{\mbox{th}}$ crop. The threshold may also be thought of as a prescribed coverage level. For a given crop, some of the declarations might coincide with yields that are above the drought threshold, $\mu_c^{(j)}$. For that particular crop, the declaration of drought can be thought of as a false declaration. This leads us to define the proportion of coincident (or `true') declarations to be
\begin{equation}
\psi_j^\tau=\frac{1}{|\Gamma|}\sum_{t\in\Gamma}H\left(\mu_c^{(j)}-\mathit{\Upsilon}_j(t)\right),
\end{equation}
where $H(\cdot)$ is the Heaviside step function and $\Gamma$ is the set of times when drought declarations were made. The proportion of declarations that are false is then given by $\psi^f=1-\psi^\tau$. The proportion $\psi^f$ gives an indication of how often farmers get financial assistance when they should not. The crop-specific drought threshold need not correspond to the frequency of drought declarations. It can be selected purely by other considerations such as the yield potential of the specific crop. If selected by other considerations, it leads us to define the crop-specific drought frequency as
\begin{equation}
\omega_j=F_j\left(\mu_c^{(j)}\right).
\end{equation}
The threshold $\mu_c^{(j)}$ can also be thought of as the pre-specified crop-specific coverage level. Given that $\lambda_j(t)$ is the price per hectorage, the crop will experience a production loss amounting to 
\begin{equation}
L_j(t)=\lambda_j(t)\max\left\{0,\mu_c^{(j)}-\mathit{\Upsilon}_j(t)\right\},
\end{equation}
which can be thought of as an indemnity payment for the specific crop. The distribution of $L_j(t)$ can be obtained from the truncated distribution of the yield, following which we can obtain the mean loss $\mathbb{E}\left[L_j(t)\right]$. The weighted loss for the cluster of crops is
\begin{align}
L_\theta(t)=\sum_{j=1}^J\theta_j(t)L_j(t),
\end{align}
where $\theta_j(t)$ is the random variable for the proportion of area in which the $j$th crop was planted. Moreover, $\sum_{j=1}^J\theta_j(t)=1$. The above model sees each individual loss as a random variable. Consequently, the weighted loss experienced by the pool of crops is also a random variable. The expectation of the weighted loss is thus given by 
\begin{align}
\mathbb{E}[L_\theta(t)]&=\sum_{j=1}^J\mathbb{E}\left[\theta_j(t)L_j(t)\right]\\
&=\sum_{j=1}^J\mathbb{E}[\theta_j(t)]\mathbb{E}[L_j(t)]+\sum_{j=1}^J\mbox{Cov}[\theta_j(t),L_j(t)]
\end{align}
The variance of the weighted loss at year $t$ is then given by
\begin{align}
\mbox{Var}(L_\theta(t))&=\mbox{Var}\left(\sum_{j=1}^J\theta_j(t)L_j(t)\right)\\
&=\sum_{i=1}^J\sum_{j= i}^J\mbox{Cov}[\theta_i(t)L_i(t),\theta_j(t)L_j(t)]\\
&=\sum_{j=1}^J\mbox{Var}\left[\theta_j(t)L_j(t)\right]+\sum_{i=1}^J\sum_{j=1,j\neq i}^J\mbox{Cov}[\theta_i(t)L_i(t),\theta_j(t)L_j(t)]\\
\nonumber
&=\sum_{j=1}^J\mbox{Cov}\left[\theta_j(t)^2,L_j(t)^2\right]+\sum_{j=1}^J\mathbb{E}\left[\theta_j(t)^2\right]\mathbb{E}\left[L_j(t)^2\right]-\\
&\sum_{j=1}^J\Big\{\mbox{Cov}\left[\theta_j(t),L_j(t)\right]+\mathbb{E}[\theta_j(t)]\mathbb{E}[L_j(t)]\Big\}^2+\sum_{i=1}^J\sum_{j=1,j\neq i}^J\mbox{Cov}[\theta_i(t)L_i(t),\theta_j(t)L_j(t)].
\end{align}
If $\theta_i$, $\theta_j$, $L_i$ and $L_j$, where $i\neq j$, are pairwise independent, then the above formula reduces to
\begin{equation}
\mbox{Var}(L_\theta(t))=\sum_{j=1}^J\mbox{Cov}\left[\theta_j(t)^2,L_j(t)^2\right].
\end{equation} 

Setting 
\begin{equation}
Z(t)=\frac{L(t)-\mathbb{E}[L_\theta(t)]}{\sqrt{\mbox{Var}[L_\theta(t)]}},
\end{equation}
we can approximate $Z(t)$ by the standard normal distribution, meaning that $Z(t)\sim N(0,1)$. In this case,
\begin{equation}
\mathbb{P}\left(\big|L_\theta(t)-\mathbb{E}[L_\theta(t)]\big|<1.96\sqrt{\mbox{Var}[L_\theta(t]}\right)\approx 0.95
\end{equation}
can be thought of as the probability that the fund set up to finance the scheme will not be exhausted. Therefore, the buffer fund needs to be set at $\mathbb{E}[L_\theta(t)]+2\sqrt{\mbox{Var}[L_\theta(t)]}$ per hectorage to maintain the probability of ruin at 2.5\%. More generally, the fund should be set to be
\begin{equation}
F_\theta=A\left(\mathbb{E}[L_\theta(t)]+\eta\sqrt{\mbox{Var}[L_\theta(t)]}\right)
\end{equation}
where $A=\sum_{j=1}^JA_j$ and $\eta$ is a parameter to be chosen to be consistent with the risk appetite of the insurer or fund manager. In essence the $\eta$ should be chosen to minimise the probability of ruin.

It is important to measure the effect of using mixed farming to mitigate risk. Whether or not mixing crops is effective as a risk mitigation strategy can be determined by assessing how the variance of $L_\theta(t)$ compares with the weighted average variance of each $L_j(t)$. The coefficient of effectiveness can be used for this. If $V$ denotes the weighted average variance of the losses, then the {\it coefficient of effectiveness} is given by
\begin{equation}
\phi_\theta=\frac{\mbox{Var}[L_\theta]}{\mathbb{E}[V]},
\end{equation}
where
\begin{equation}
V=\sum_{j=1}^J\theta_j\mbox{Var}[L_j].
\end{equation}
Setting $\mathbb{E}[\theta_j]=\alpha_j$, it turns out that
\begin{equation}
\mathbb{E}[V]=\sum_{j=1}^J\alpha_j\mbox{Var}[L_j],
\end{equation}
from whence the coefficient of effectiveness becomes
\begin{equation}
\phi_\theta=\frac{\mbox{Var}[L_\theta]}{\sum_{j=1}^J\alpha_j\mbox{Var}[L_j]}.
\end{equation}
The lower the value of $\phi_\theta$, the more effective the risk pooling approach. If $\theta_j$ is constant for all $j$ with $\theta_j=\alpha_j$, then the above formula reduces to
\begin{align}
\phi_\theta &=\frac{\left(\sum_{j=1}^J\alpha_j(t)^2\mbox{Var}\left[L_j(t)\right]+\sum_{i=1}^J\sum_{j\neq i}^J\alpha_i(t)\alpha_j(t)\mbox{Cov}\left(L_i(t),L_j(t)\right)\right)}{\left(\sum_{j=1}^J\alpha_j\mbox{Var}\left[L_j(t)\right]\right)}.
\end{align}
If the losses are independent, in which case $\mbox{Cov}\left(L_i(t),L_j(t)\right)=0$ whenever $i\neq j$, then the coefficient of effectiveness becomes
\begin{equation}
\phi_\theta=\left(\sum_{j=1}^J\alpha_j(t)^2\mbox{Var}\left[L_j(t)\right]\right)\Bigg/\left(\sum_{j=1}^J\alpha_j(t)\mbox{Var}\left[L_j(t)\right]\right).
\end{equation}
In addition, if the variances are equal, then the formula for the coefficient of effectiveness becomes
\begin{equation}
\phi_\theta=\sum_{j=1}^J\alpha_j(t)^2.
\end{equation}
Consequently, using the method of Lagrangian multipliers under uncorrelated risk, it follows that $\phi_\theta$ attains a minimum when $\alpha_i=\alpha_j$ for all $i\neq j$, in which case the minimum is $\phi_\theta=1/J$.

The gain made from the $j^{\mbox{th}}$ crop in the year $t$, denoted by $G_j(t)$, can be defined via
\begin{equation}
G_j(t)=\lambda_j(t)\max\left\{0,\mathit{\Upsilon}_j(t)-\mu_c^{(j)}\right\}.
\end{equation}
The total gain during a given season is
\begin{equation}
G_\theta(t)=\sum_{j=1}^J\theta_jG_j(t)
\end{equation}
A payout to farmers should be made as long as $L_\theta(t)>0$, in which case the losses are non-zero. Otherwise there should be no payout. The surplus for the $j^{\mbox{th}}$ crop in a given planting season is then given by
\begin{align}
S_j(t)&=G_j(t)-L_j(t).
\end{align}
If $\mathbb{E}[S_j(t)]\le 0$, then the $j^{\mbox{th}}$ crop is not insurable and should be removed from the cluster of crops that are covered. In essence the crop makes no business sense. Without loss of generality, we assume that each crop in the cluster satisfies the condition that $\mathbb{E}[S_j(t)]>0$, implying that it makes business sense. The weighted surplus per hectorage is
\begin{align}
S_\theta(t)&=\sum_{j=1}^J\theta_j(t)S_j(t).
\end{align}
The expected surplus per hectorage is
\begin{equation}
\mathbb{E}[S_\theta(t)]=\sum_{j=1}^J\mathbb{E}[\theta_j(t)S_j(t)].
\end{equation}
If $\mathbb{E}[S_\theta(t)]>0$, then the cluster of crops makes business  sense and may be insurable. Note, however, that the condition that $\mathbb{E}[S_j(t)]>0$ does not guarantee that $\mathbb{E}[S_\theta(t)]>0$. In order to be given a loan and/or be allowed to take out insurance, it is necessary to establish that the average surplus per hectorage is positive, i.e. $\mathbb{E}[S_\theta(t)]>0$. In the case that the proportion of hectorage planted is constant, i.e. $\theta_j(t)=\alpha_j$, then
\begin{align}
\mathbb{E}[S_\theta(t)]=\sum_{j=1}^J\alpha_j\mathbb{E}[S_j(t)],
\end{align}
where the linear property of the expectation operator has been used.  Similarly, if $\theta_j(t)$ and $S_j(t)$ are independent for all $j$, then
\begin{align}
\mathbb{E}[S_\theta(t)]&=\sum_{j=1}^J\mathbb{E}[\theta_j(t)]\mathbb{E}[S_j(t)].
\end{align}
Since the proportion of hectorage planted for each crop cannot be negative, it follows that $\mathbb{E}[\theta_j(t)]\ge0$. Consequently, under the foregoing independence condition, it is guaranteed that expected surplus per hectorage will be non-negative, i.e. $\mathbb{E}[S_\theta(t)]\ge0$, which implies that the cluster of crops makes business sense and is thus insurable. 

Each crop-specific threshold may be chosen to be the minimum required for profitability, having taken into account production costs (including all seasonal inputs) for a given season. In the instance of self-insurance, one would use the surplus to take care of the years of want. In the absence of annual loan instalment obligations, self-insurance is possible if and only if $\mathbb{E}[S_\theta(t)]\ge0$. In this case, the farming is profitable. Given that the loan instalment per area planted per season is $l$, it is important to have $\mathbb{E}[S_\theta(t)]>l$ to be able to use the proceeds of farming to service the instalments. Without insurance, the amount left after paying instalments is 
\begin{equation}
R=\mathbb{E}[S_\theta(t)]-l.
\end{equation}
If the frequency of drought declarations is $\omega$, then the average amount paid towards instalments per season is
\begin{equation}
l_1=(1-\omega)l+p\omega l,
\end{equation}
where $(1-p)$ is the insured benefit level, with $0\le p\le 1$. Denoting the premium rate by $\gamma_\theta$, the amount of premium paid to take out insurance is $\gamma_\theta l$, the portion of surplus left after making payments towards instalment and premiums is
\begin{equation}
R_1=\mathbb{E}[S_\theta(t)]-(1-\omega)l-p\omega l-\gamma_\theta l.
\end{equation}
Note that $\omega$ can be chosen to be consistent with a specific collection of crops. The premium contributions and subsidies may be set to reflect both the frequency and severity of the drought events. In order for the farmer to remain solvent, it is important to have $R_1>0$. This condition translates to the inequality
\begin{equation}
(\gamma_\theta+p\omega)<\frac{\mathbb{E}[S_\theta(t)]}{l}-(1-\omega).
\label{eqn:cond}
\end{equation}
An actuarially sound premium level should satisfy the equation 
\begin{equation}
\gamma_\theta=\omega(1-p).
\label{eqn:sound}
\end{equation} 
When premiums are actuarially fair, inequality~(\ref{eqn:cond}) reduces to the requirement that $\mathbb{E}[S_\theta(t)]>l$. What happens if actuarially sound premiums lead to insolvency, i.e. $R_1\le0$ or equivalently $\mathbb{E}[S_\theta(t)]\le l$? In this case, under the condition that $\mathbb{E}[S_\theta]\ge0$, there is justification for Government subsidy and the proportion of the rate that cushions the farmer against insolvency is
\begin{equation}
\nu \ge 1-\frac{\mathbb{E}[S_\theta(t)]}{l}.
\end{equation}
The farmer's premium rate should then be set to 
\begin{equation}
\gamma_\theta=\omega(1-p)(1-\nu).
\label{eqn:prem}
\end{equation}
If $\mathbb{E}[S_\theta]<0$, then this is a case for full Government subsidy and Government should pay the full premium rate. The formulae for premium rate and Government subsidy are respectively given by
\begin{equation}
 \gamma_\theta=\left\{\begin{array}{cl}
                       0 & \mbox{if} \quad \mathbb{E}[S_\theta]<0,\\
                       \omega(1-p)(1-\nu) & \mbox{if}\quad\mathbb{E}[S_\theta] \in[0,l)\\
                       \omega(1-p) & \mbox{if} \quad \mathbb{E}[S_\theta]\ge l
                      \end{array}
\right.
\label{eqn:gammq}
\end{equation}
and
\begin{equation}
 \kappa_\theta=\left\{\begin{array}{cl}
                       \omega(1-p) & \mbox{if} \quad \mathbb{E}[S_\theta]<0,\\
                       \omega(1-p)\nu & \mbox{if}\quad \mathbb{E}[S_\theta] \in[0, l)\\
                       0 & \mbox{if} \quad \mathbb{E}[S_\theta]\ge l.
                      \end{array}
\right.
\label{eqn:kq}
\end{equation}
where $\nu\in\left[1-\frac{\mathbb{E}[S_\theta]}{l},1\right]$. It is possible to deduce that $\gamma_\theta+\kappa_\theta=\omega(1-p)$ for all  $\mathbb{E}[S_\theta]$. In the next section, we select $\nu=1-\mathbb{E}[S_\theta]/l$ whenever $\mathbb{E}[S_\theta]\in[0,l)$. In applications, Government is free to select greater subsidy rates within the range according to what it can afford. In the special case when $\theta_j=1/J$, then the corresponding variables and the parameters become $\phi$, $L(t)$, $G(t)$, $\gamma$, $\kappa$, $S(t)$, etc. 
\section{Numerical Analysis}
In this section, we consider the potential benefit of drought management through planting multiple field crops, applying the theory from the previous section. This approach is a form of risk pooling strategy. In order to apply the theory from the previous section, we consider three field crops, Sorghum, Maize and Pulses that are planted in Botswana. The data for these crops dates from 1979 to 2003 and was obtained from the Ministry of Agriculture. This study considers commercial farming and focuses on the variables, area planted in hectares and crop yield in terms of Kilograms per hectare. In order to plant these crops, farmers can take loans from participating lending institutions. The loans are subject to annual instalments. When they take the loans, they are also required to take out insurance offered through the agricultural credit guarantee scheme. 

In order to benefit from the scheme, each farmer pays 5\% of the instalment towards an insurance premium whilst the lending institution contributes 5\%. In total, 10\% is paid as an insurance premium. An analysis of claims data from the Ministry of Finance and Economic Development indicates that payouts from the scheme were made 2/3 of the times, which implies that the probability of drought is $\omega=2/3$. The contribution of 10\% entitles the insured (who is the farmer) to a cover of up to 85\% of his annual instalment if a drought is declared. According to Equation~(\ref{eqn:sound}), a corresponding actuarially sound premium rate is 56.7\% of the instalment. The premium rate is actuarially sound as long as the insured benefit level is exactly 85\% and not less. Equation~(\ref{eqn:sound}) applies when the insured benefit level is exactly $(1-p)$ and overall operations are not conducted at a loss.

\begin{figure}[!t]
%\centering
\hbox{
\includegraphics[height=6.0cm,width=6.5cm]{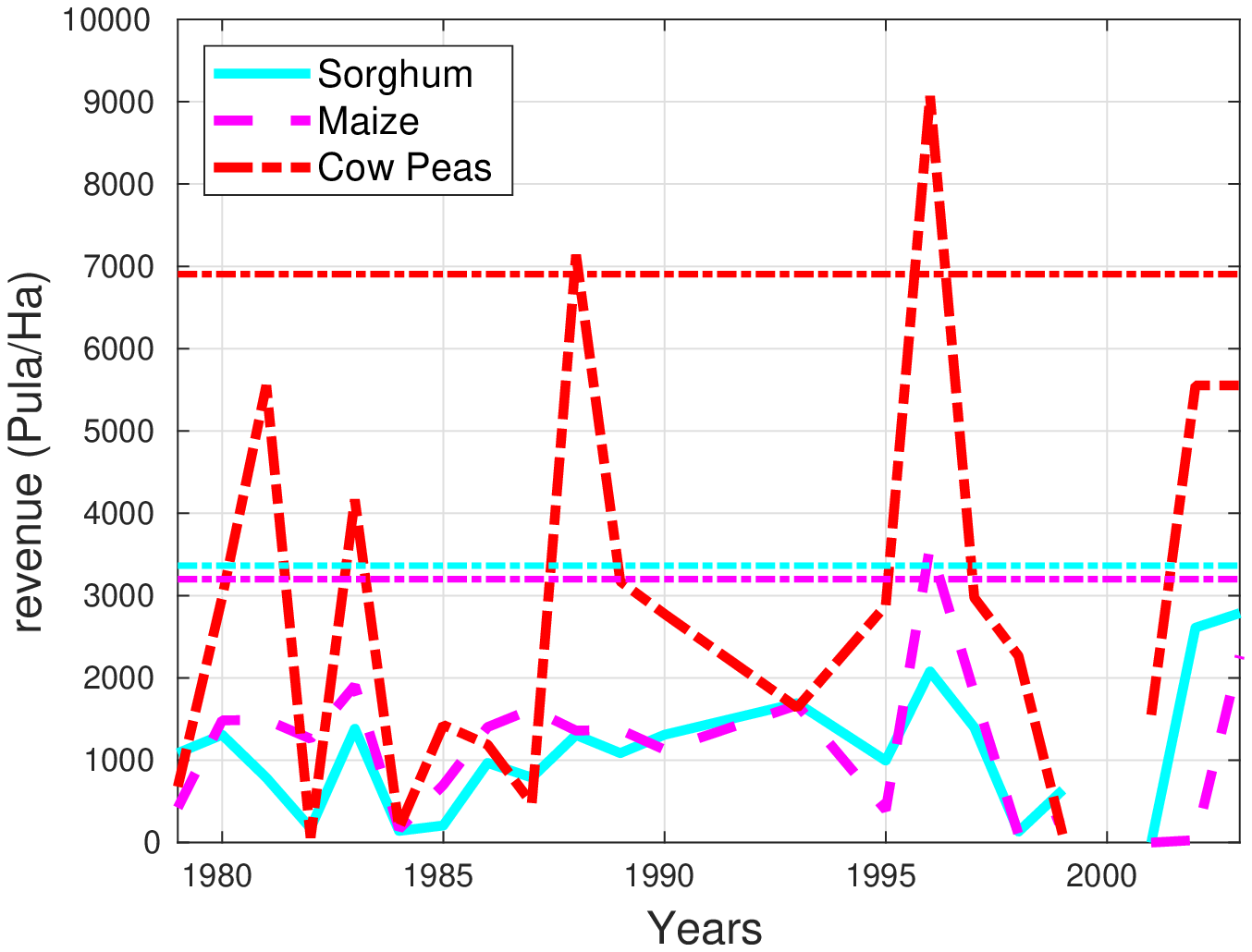},
\includegraphics[height=6.0cm,width=6.5cm]{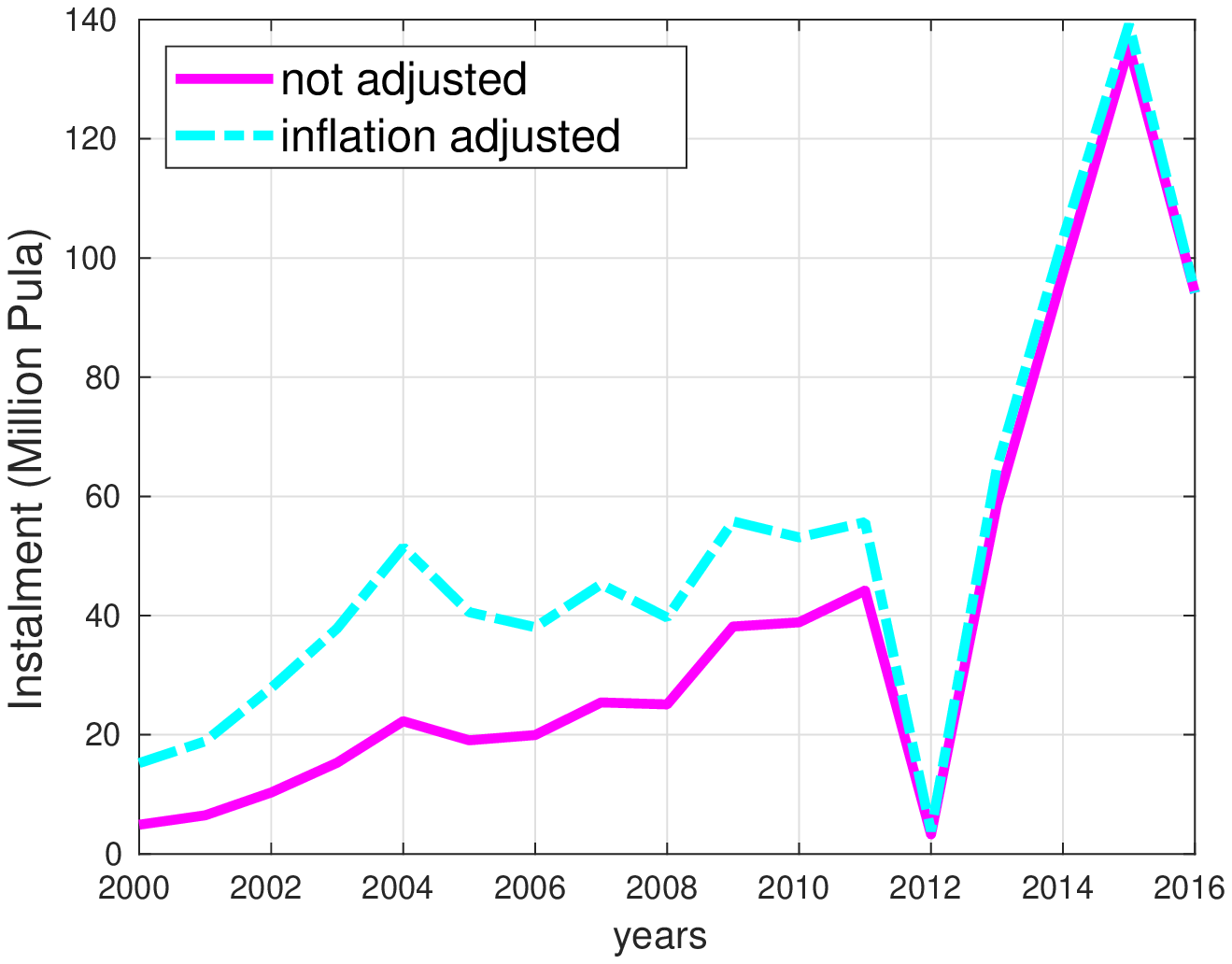}
}
\caption[Map Graph]{\parbox[t]{0.9\linewidth}{\em\small Graphs showing the (left) revenue per hectare for the three field crops, sorghum, maize and cow-peas and (right) the total instalments paid towards arable farming loans. The straight lines on the left graph correspond to the input costs for the three field crops.}}
\label{fig:yield}
\end{figure}

Since insurance requires an understanding of the revenue that proceeds from these crops, yield time series for these crops can be multiplied by costs per kilogram to obtain revenue per hectare. Using the variables from the previous sections, we want to obtain a time series plot of $\lambda_j(t)\mathit{\Upsilon}_j(t)$, where $\Upsilon_j$ and $\lambda_j$ are the yield and price per hectare respectively, for the $j$th crop in the cluster. In order to obtain the variations in revenue over time, having taken inflation adjustments into account, 2016/2017 prices from Botswana Agricultural Marketing Board (BAMB) were used. Those prices were, P1.75, P1.70 and P11.90 for maize, sorghum and cow peas respectively. Using those values allowed comparison with estimates of farm input costs provided by the Department of Crop Production. The graphs for Maize, Sorghum and Cow peas are shown in Figure~\ref{fig:yield}. The three straight lines indicate the corresponding estimates of farm inputs provided by the Department of Crop Production. These costs have been heavily subsidised under ISPAAD, Integrated Support Programme for Arable Agricultural Development. The colour of each straight line matches that of the corresponding crop. According to these graphs, the revenue for each crop is always (except once for maize and cow peas) lower than the cost of inputs. This implies that the farming of the respective crops is generally operated at a loss. If these estimates of farm inputs are accurate, then the farming of these field crops requires full subsidy every year in order to be sustained. In that case, giving partial financial assistance only during the years of national drought declarations cannot suffice. Since farmers have remained in business under the current arrangement of the Agricultural Credit Guarantee Scheme, it is highly likely that the yield figures reported by the commercial farmers are grossly inaccurate. The under reporting can happen especially when the commercial farmers hope to attract more subsidy from Government.

\begin{figure}[!t]
%\centering
\hbox{
\includegraphics[height=6.0cm,width=6.5cm]{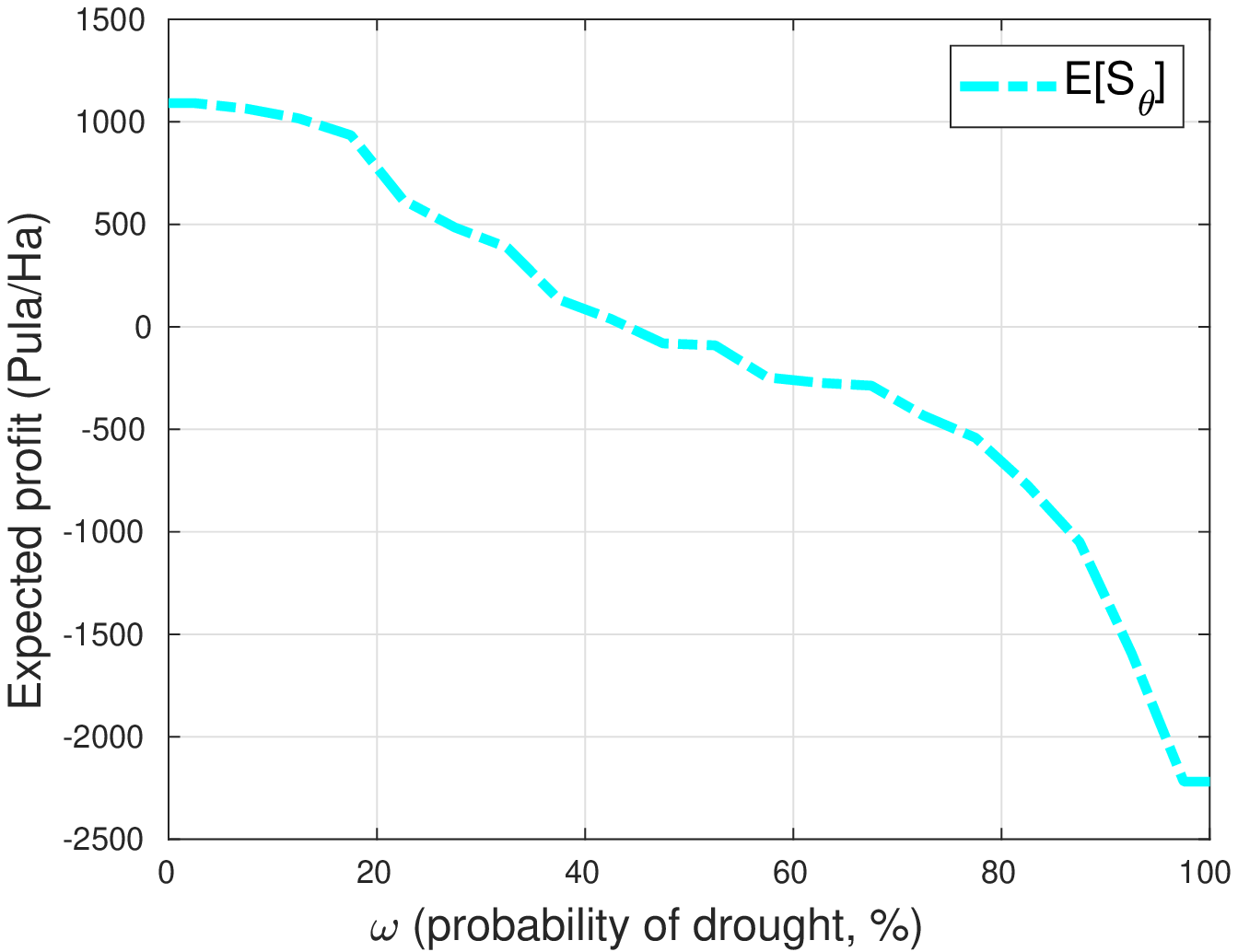},
\includegraphics[height=6.0cm,width=6.5cm]{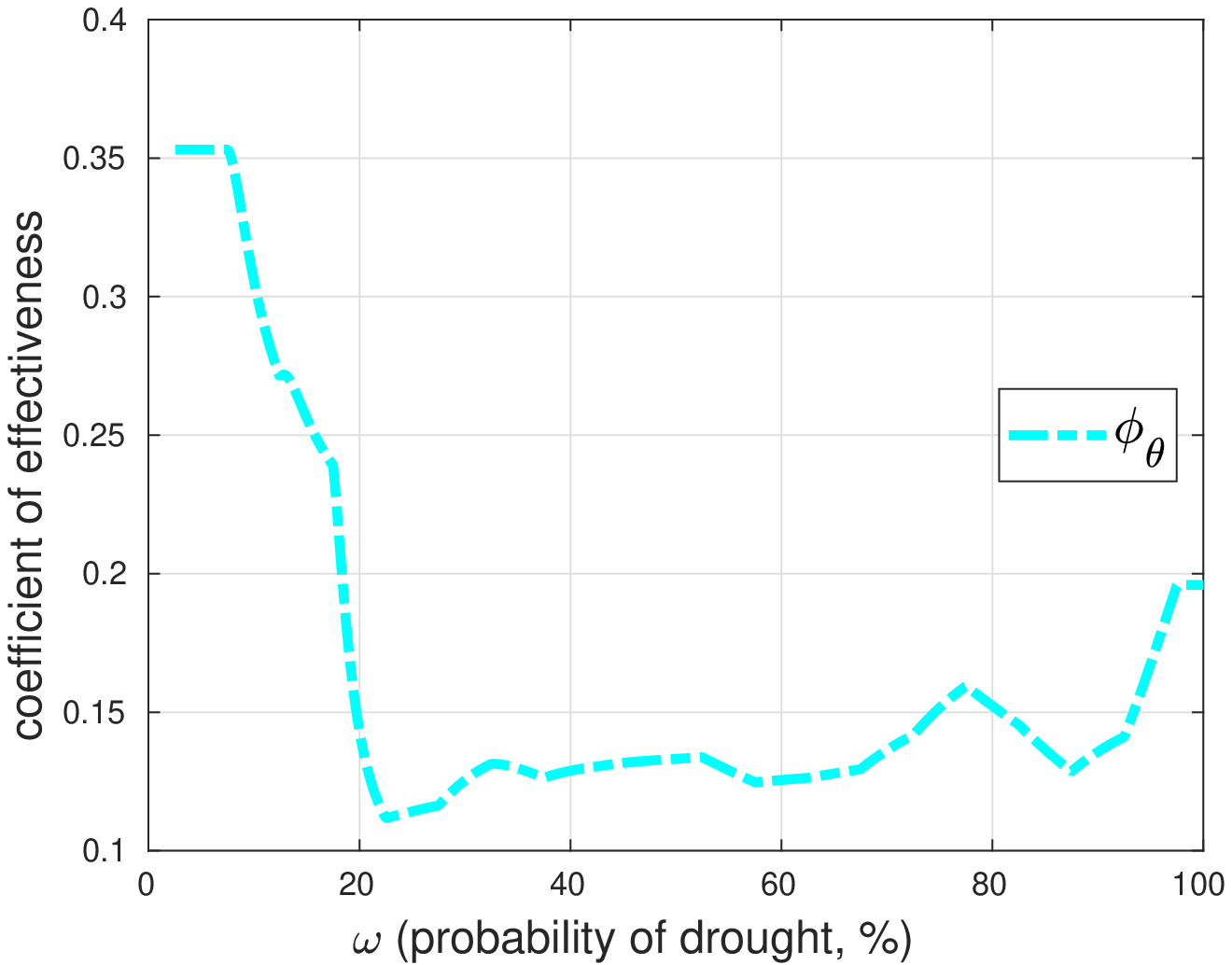}
}
\caption[Map Graph]{\parbox[t]{0.9\linewidth}{\em\small Graphs showing the (left) expected profit per hectare for the three field crops, sorghum, maize and cow-peas and (right) the coefficient of effectiveness.}}
\label{fig:coeff}
\end{figure}

An alternative way of setting up a loss threshold is to use the percentage of drought declarations as discussed in the previous section. Using data for commercial farming at a national level, graphs of the profit obtained under varying values of probability of drought are shown in Figure~\ref{fig:coeff}. Figure~\ref{fig:coeff} shows the expected profit $\mathbb{E}[S_\theta]$ and the coefficient of effectiveness $\phi_\theta$, each as a function of the probability of drought, $\omega$. $\mathbb{E}[S_\theta]$ is the expectation of the surplus revenue due to mixing the field crops. On the right hand of the figure are graphs of the coefficient of effectiveness for the surplus revenue per square area. On the graphs, $\phi_\theta$ is the coefficient of effectiveness resulting from mixing up the crops. Recall that the lower the coefficient of effectiveness, the more effective the risk pooling approach. According to these graphs, mixing field crops is an effective risk pooling strategy as the probability of drought increases. In what sense is mixing of field crops effective? It leads to lower variations in the fund due to claims, thus reducing the probability of ruin.
\begin{figure}[!t]
\centering
\includegraphics[height=7.5cm,width=8.5cm]{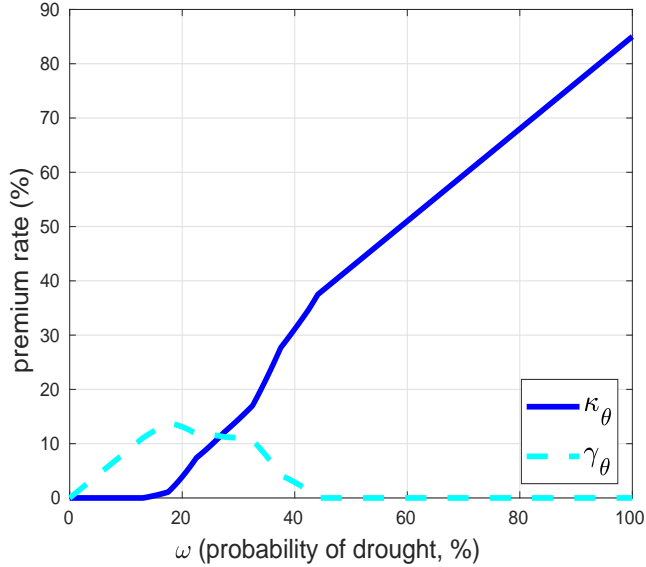}
\caption[Map Graph]{\parbox[t]{0.9\linewidth}{\em\small Graphs of premium rate versus probability of drought. In the graphs, $\gamma_\theta$ is the premium rate that the farmer should pay subject to a government subsidy $\kappa_\theta$.}}
\label{fig:premium}
\end{figure}

To assess what premium rate should be paid by farmers subject to Government subsidy, one should consider Figure~\ref{fig:premium}. The nonlinear formulae upon which the graphs in the figure are based are given in Equations~(\ref{eqn:gammq}) and~(\ref{eqn:kq}). The loan instalment per area planted was computed based on an average of instalments paid and area planted for the past 10 years. The idea was to use a moving average of instalments to account for climate change. The choice of the length of time over which to compute the average annual instalment per hectare is based on no hard rule, except that we suggest using a time period that is less than climate time scales. In producing the graphs, the value of instalments used was $l=P1008/\mbox{Ha}$. Here, the farmer's premium rate is $\gamma_\theta$ whilst the Government subsidy is $\kappa_\theta$, where the subscript $\theta$ denotes that the  expectations were weighted by the area planted for each field crop. According to the graphs in Figure~\ref{fig:premium}, Government subsidy is required when a threshold of 17\% probability of drought is crossed. From then on, the subsidy rate required increases nonlinear until the drought frequency exceeds 43\%, at which drought frequency full subsidy is required. These findings suggest that the current rate of drought declarations merits full subsidy on seasonal loans, which is doubtful because the farmers have not gone burst without it. 
\section{Discussion}
This paper sought to investigate the possibility for Government of Botswana to setup a fund to finance annual instalments for dry-land arable farming in times of drought. It notes that, at the moment, there is no fund setup to assist farmers who have taken loans with participating banks. If available, loss data can be used to determine what the size of the fund should be. The fund should be setup to minimise the probability of ruin by an aggregate loss that is large enough to do so. In this paper, it is suggested that yield data for insurable crops provides a good proxy for loss data. Whilst there is no one way for setting up the threshold that determines losses and gains, for a start, past records of drought declarations could be used to set the thresholds. Each crop would then have a threshold consistent with the probability of drought. The thresholds can then be used to determine the financial losses (or gains) corresponding to the yield shortfalls (or excesses). The fund should then be setup to be equivalent to the area weighted losses for the several crops within an insured cluster. The concept of using the area weighted losses is supported by results of the empirical example considered in this paper. The said fund is meant to assist farmers who have taken out loans with participating banks to pay instalments in years of poor yield or crop failure. If a drought has been declared, their instalments should be paid off through funds from the scheme. This paper argues that these funds need not and should not be sought in an adhoc manner.

It should be understood that in order for farmers to benefit from the scheme, they should each pay a premium that is actuarially sound. The current status quo is that all farmers pay a flat rate regardless of what crops they are growing. This paper argues that a more prudent approach is to set rates that are crop specific and it provided an algorithm that also determines an appropriate Government subsidy for each cluster of crops under consideration. As an empirical example, a cluster of field crops comprising Maize, Sorghum and Cow Peas was considered. It was found that when the probability of drought exceeds 43\%, yield data at a commercial scale indicates a lack of profitability. This finding is in conflict with the fact that farmers still remain in business when, according to national declarations, the climatological probability of drought is about 2/3 ($\sim 67\%$). An actuarially sound premium rate consistent with this probability of drought at an insured benefit level of 85\% is 56\%, but the contributions from both farmer and the lending institution amounts to only 10\%. According to empirical results presented in this paper, this subsidy rate still wouldn't be sufficient to keep the farmers in business because it is less than a full subsidy. This begs the question of why commercial farmers are still in business. Where is the catch? There are two possibilities: Either the blanket national drought declarations are heavily flawed in relation to dry land arable farming or the commercial farmers grossly under report their crop yields. In light of these possibilities, there is need for a thorough review of the applicability of national drought declarations to arable farming.

It is important to note that, whereas the payment of a premium of 10\% entitles the farmer to a variable benefit level of up to 85\%, the reality is that Government has invariably paid out a fixed rate of 85\%. There is, however, still a need to vary the benefit level according to severity. The effect of varying the insured benefit level with severity is a subject for further consideration. To avoid the problem of moral hazard and asymmetrical information, remote sensing information may be used to assess severity so as to determine an appropriate payout. This will be an important question to address in further research. Whilst the use of indices has been strongly criticised for its failures due to basis risk~\citep{reeves-17},~\cite{eze-2020} suggested that basis risk can be reduced by calibrating the indices according to local areas where specific farms are located.
\section{Conclusions}
This paper argued for mixing field crops as a drought mitigation strategy and presented a methodology that is applicable to any chosen cluster of dry-land crops. It presented a way for Government to setup a drought fund for dry-land arable farming. Nonlinear formulae for setting up premium rates and subsidies were presented and tested with a numerical example. The formulae provide thresholds for Government subsidy rates and indicate when a full subsidy is warranted. In here was also presented empirical evidence that national drought declarations are not consistent with yield data for the main field crops. If current drought declaration mechanisms are appropriate for dry land arable farming of field crops, then full Government subsidy is warranted. The use of area-specific, satellite based, drought indices should be explored as a way of making drought declarations. Should these be used, care should be taken to minimise basis risk and provide early compensation or assistance to affected farmers. 

The drought risk management approach presented relies solely on climatological information to make farming decisions and set up a subsidy fund. Moving forward, it will be interesting to consider using seasonal climate forecasts to decide whether or not to farm and determine appropriate quantities to plant. If the forecasts are more skillful than the climatological distribution, then the losses can be reduced. Consequently, required Government subsidies can be lowered. There are already some developments on the use of seasonal forecasts. For instance,~\cite{yi-2020} discuss the use of ENSO forecasts for rate making. Further work should consider the use of country or region-specific Seasonal Climate Forecasts by different stakeholders, including farmers and Governments.
\section*{Acknowledgements}
Thanks to Segomotso Sabone, Benjamin Goemekgabo, Bright Ramaina, Nnyaladzi Batisani and Kebitsang Powe for providing useful data and engaging in discussions that helped put this document together. This work was financially supported by Botswana Government.
\bibliographystyle{jofbib}
\bibliography{refs}
\end{document}